\documentclass[graybox]{svmult}

\usepackage{type1cm}        
\usepackage{makeidx}         
\usepackage{graphicx}        
\usepackage{multicol}        
\usepackage[bottom]{footmisc}

\usepackage{newtxtext}       %
\usepackage{newtxmath}       
\usepackage{amsfonts}
\usepackage{amsmath}

\makeindex             

\begin{document}

\title*{Artificial Intelligence-Based Image Reconstruction in Cardiac Magnetic Resonance}
\author{Chen Qin and Daniel Rueckert}
\institute{Chen Qin \at 
Department of Electrical and Electronic Engineering, Imperial College London, London, UK \\
Institute for Digital Communications, School of Engineering, University of Edinburgh, UK\\
\email{c.qin15@imperial.ac.uk}
\and Daniel Rueckert \at Department of Computing, Imperial College London, London, UK \\ Klinikum rechts der Isar, Technical University of Munich, Munich, Germany\\ \email{d.rueckert@imperial.ac.uk}}
%
%
\maketitle


\abstract{Artificial intelligence (AI) and Machine Learning (ML) have shown great potential in improving the medical imaging workflow, from image acquisition and reconstruction to disease diagnosis and treatment. Particularly, in recent years, there has been a significant growth in the use of AI and ML algorithms, especially Deep Learning (DL) based methods, for medical image reconstruction. DL techniques have shown to be competitive and often superior over conventional reconstruction methods in terms of both reconstruction quality and computational efficiency. The use of DL-based image reconstruction also provides promising opportunities to transform the way cardiac images are acquired and reconstructed. In this chapter, we will review recent advances in DL-based reconstruction techniques for cardiac imaging, with emphasis on cardiac magnetic resonance (CMR) image reconstruction. We mainly focus on supervised DL methods for the application, including image post-processing techniques, model-driven approaches and $k$-space based methods. Current limitations, challenges and future opportunities of DL for cardiac image reconstruction are also discussed.}


\section{Introduction}
\label{sec:1}
Artificial intelligence (AI) and Machine Learning (ML), and in particular Deep Learning (DL), have been active and fast-moving areas of research in the field of medical imaging. These techniques have great potential for transforming  clinical practice in the coming years \cite{pesapane2018artificial}. DL-based methods have also led to significant improvement across various applications in cardiac imaging, such as cardiac image segmentation \cite{chen2020deep,bai2018automated}, myocardial motion tracking \cite{duchateau2020machine,qin2018joint,qin2020biomechanics} and image enhancement \cite{oktay2017anatomically,oktay2016multi}. Cardiac image reconstruction, as one of the key step in cardiac imaging workflow, has also witnessed the important growth of the use of DL algorithms, though the development of which is relatively new in comparison to other cardiac imaging applications.

In cardiac imaging, cardiac magnetic resonance (CMR) imaging is one of the most useful and important imaging modalities for the non-invasive assessment of morphology and function of the heart. It provides images with excellent soft tissue contrast as well as high spatial and temporal resolution and is therefore widely used for the assessment of cardiovascular diseases. However, the acquisition speed of CMR is limited, which is on account of the need of obtaining images of high temporal and spatial resolution, as well as the cardiac and respiratory-induced motion that needs to be taken into consideration during the acquisition process \cite{bustin2020compressed}. 

To improve and accelerate the CMR acquisition process, several major advances have been made over the last few decades. Parallel imaging (PI) \cite{pruessmann1999sense,griswold2002generalized,sodickson1997simultaneous} is one of the techniques frequently used to accelerate the MR acquisition. In PI, the data is acquired using an array of multiple independent receiver coils, allowing for a reduction in the number of phase-encoding steps and enabling the combination of undersampled data from each receiver coil given the additional coil information. Particularly, for dynamic CMR imaging where substantial correlations in $k$-space and time exist, most PI-based imaging strategies have been designed to acquire part of the desired $k$-$t$ measurements and then reconstruct images by exploiting spatio-temporal redundancies in combination with coil sensitivity information \cite{kellman2001adaptive,breuer2005dynamic,huang2005k}. Undersampled image reconstruction approaches such as compressed sensing (CS) have also been proposed for CMR image reconstruction. CS-based methods \cite{lustig2007sparse,ye2019compressed} exploit signal sparsity with some sparse transforms such as finite differences or wavelets operators, and recover original image from undersampled $k$-space data using nonlinear reconstructions. For dynamic CMR image reconstruction, a common practice for CS-based approaches is effectively to exploit the spatio-temporal redundancies via enforcing the signal sparsity in spatio-temporal or its transform domain against the consistency of the undersampled acquired data, such as some state-of-the-art $k$-$t$ methods \cite{lustig2006kt,jung2007improved,otazo2010combination}. Combination of CS with low-rank schemes \cite{lingala2011accelerated,otazo2015low}, spatio-temporal partial separability \cite{zhao2012image} and patch-based regularisation framework \cite{yoon2014motion,mohsin2017accelerated} have also been proposed to exploit the spatio-temporal correlations or geometric similarities for the dynamic CMR image reconstruction.

Though CS-based methods have shown great potential for accelerated CMR imaging, they often impose strong assumptions on the data and require nontrivial manual adjustments of hyperparameters depending on the application. In addition, reconstruction speed of these methods is often slow due to the iterative optimisation used, and in the context of dynamic CMR imaging, the additional time domain further increases the computational demand. To enable progress beyond the limitations of conventional CS-based reconstruction methods, DL approaches have been proposed for CMR image reconstruction. These DL solutions can be roughly categorised as image post-processing techniques, model-driven image reconstruction methods or $k$-space interpolation approaches. Most of these approaches typically rely on large databases of MR scans as training data to exploit the prior knowledge. DL approaches have shown great performance in achieving high image reconstruction quality with extremely fast reconstruction speed, which is promising to further accelerate CMR acquisition as well as to improve the reconstruction.

In this chapter, we will first briefly introduce the problem formulation of CMR image reconstruction as an inverse problem, and then we will discuss the recent advances in DL for CMR image reconstruction with a major focus on supervised learning. Finally, current limitations, challenges and opportunities of DL for CMR reconstruction will be discussed.

\section{Problem Formulation for Image Reconstruction}
\label{sec:2}
A general forward model for the MR imaging process can be formulated as 
\begin{equation}
    \mathbf{y} = \mathbf{Ex} + \mathbf{n}, \label{eq:ip}
\end{equation}
where $\mathbf{x} \in \mathbb{C}^{N_x}$ denotes the complex-valued MR images to be recovered, $\mathbf{y} \in \mathbb{C}^{N_y}$ represents the undersampled $k$-space measurements corrupted by additive Gaussian noise $\mathbf{n} \in \mathbb{C}^{N_y}$, and $\mathbf{E}:\mathbb{C}^{N_x}\to\mathbb{C}^{N_y}$ is a linear forward operator which describes the encoding process modelling the MR physics. Here $N_x$ and $N_y$ define the dimensions of the reconstruction $\mathbf{x}$ and the $k$-space data $\mathbf{y}$ ($N_y \ll N_x$ in the undersampling scenario) and $\mathbf{E}$ includes the coil sensitivity maps (if available), Fourier transform as well as the undersampling mask according to the underlying multi-coil or single-coil setting. In dynamic CMR imaging, the measurements $\mathbf{y}$ are acquired in $k$-$t$ space with an additional temporal dimension $N_\textsc{T}$, and $\mathbf{x}$ is the reconstructed image sequence.

Reconstructing image $\mathbf{x}$ from measurements $\mathbf{y}$ is an ill-posed inverse problem. To ensure that there exists a solution that is unique and plausible, the reconstruction problem is commonly formulated as a regularised optimisation problem of the form:
\begin{equation}
  \label{eq1:problem_fomulation}
\hat{\mathbf{x}}={\text{argmin}}_\mathbf{x} \ \mathcal{R}(\bf{x}) + \lambda \| \bf{y} - \mathbf{E} \mathbf{x} \|^2_2.
\end{equation}
Here $\mathcal{R}$ expresses the regularisation terms that act on $\mathbf{x}$ and $\hat{\mathbf{x}}$ is recovered via enforcing the regularisation against consistency with the acquired undersampled $k$-space data. For CS-based approaches, the regularisation terms $\mathcal{R}$ are often employed in some specific transform domain with sparse representations: Common choices are wavelet transforms \cite{lustig2007sparse} or total variation \cite{block2007undersampled}. For low-rank based methods, global or local low-rankness of the dynamic images is exploited via regularising the rank or nuclear norm of $\mathbf{x}$ to estimate the missing samples. Here, the model weight $\lambda$ allows the adjustment of data fidelity based on the noise level of the acquired measurements $\mathbf{y}$. 

To solve the inverse problem formulated as Eq. \ref{eq1:problem_fomulation}, various optimisation schemes can be employed \cite{ye2019compressed}, such as alternating directional method of multiplier (ADMM) \cite{wang2008new}, split Bregman iteration \cite{goldstein2009split} and primal-dual algorithm \cite{chambolle2011first}. A common practice to optimise Eq. \ref{eq1:problem_fomulation} is to use variable splitting techniques \cite{ramani2010parallel}, where an auxiliary variable $\mathbf{z}$ with the constraint $\mathbf{z}=\mathbf{x}$ is introduced to decouple the regularisation term from the data fidelity term. This reformulates Eq. \ref{eq1:problem_fomulation} to minimize the following cost function via the penalty method:
\begin{equation}
\label{eq: penalty_function}
{\text{argmin}}_{\mathbf{x},\mathbf{z}}\  \mathcal{R}(\mathbf{z}) + \lambda \| \mathbf{y} - \mathbf{E} \mathbf{x} \|_2^2 + \mu \|\mathbf{x}-\mathbf{z}\|^2_2,
\end{equation}
Here $\mu$ is a penalty parameter. By applying alternating minimisation over $\mathbf{x}$ and $\mathbf{z}$, Eq. \ref{eq: penalty_function} can be solved via the following iterative procedures:
\begin{subequations}\label{eq:alternate_minimisation} 
\begin{align}
\mathbf{z}^{(i+1)} & = {\text{argmin}}_{\mathbf{z}} \   \mathcal{R}(\mathbf{z}) + \mu \| \mathbf{x}^{(i)} - \mathbf{z} \|^2_2  \label{eq:proximal_operator}
\\ 
\mathbf{x}^{(i+1)} & = {\text{argmin}}_{\mathbf{x}} \  \lambda \| \mathbf{y} - \mathbf{E} \mathbf{x} \|^2_2 + \mu \| \mathbf{x} - \mathbf{z}^{(i+1)} \|^2_2, \label{eq:data_fidelity}
\end{align}
\end{subequations}
where $\mathbf{x}^{(0)}$ is commonly a zero-filled reconstruction used as an initialisation and $\mathbf{z}$ can be seen as an intermediate state of the optimisation process. For MRI reconstruction, Eq. \ref{eq:data_fidelity} is often regarded as a data consistency (DC) step.


\section{Deep learning for CMR Image Reconstruction}
\label{sec:3}
\begin{figure*}[!t]
\centering
\includegraphics[width=\linewidth]{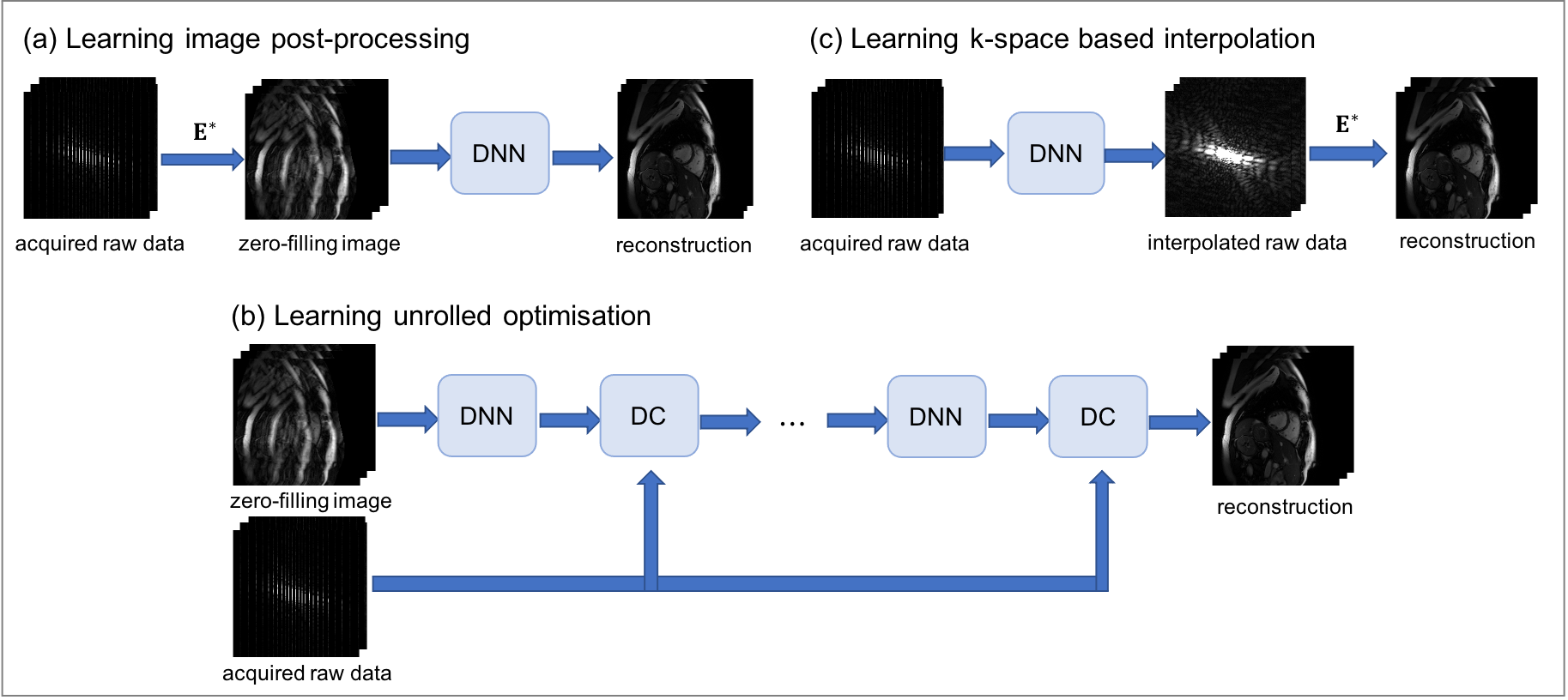}
\caption{Illustration of different types of DL-based approaches for CMR image reconstruction.
(a) Learning image post-processing for CMR image reconstruction. (b) Learning unrolled optimisation for model-driven iterative reconstruction. (c) DL for k-space based CMR reconstruction. DNN: deep neural networks, DC: data consistency.}
\label{fig:framework}
\end{figure*}

Deep learning approaches have gained in popularity for MRI reconstruction in recent years, due to their excellent capabilities in reconstructing high-quality images at extremely fast reconstruction speeds. For CMR image reconstruction, most of the DL methods can be generally categorised into three main groups: (1) image reconstruction via DL-based image post-processing; (2) model-driven unrolled DL for iterative reconstruction; and (3) $k$-space based interpolation. Schematic illustrations of these DL techniques for CMR image reconstruction are shown in Fig. \ref{fig:framework}.

\subsection{CMR Reconstruction via DL-based Image Post-processing}
The first group of methods for CMR reconstruction relies on image post-processing techniques for enhancement in the image domain. These approaches learn an end-to-end mapping from the zero-filled reconstructions to the fully-sampled reference image without any requirement for prior knowledge about the image acquisition process. 
For instance, U-net architectures are commonly used for the CMR image post-processing to reduce noise-like image artefacts \cite{hauptmann2019real,kofler2019spatio}. 
They receive inputs in form of reconstructed, undersampled CMR sequences and learn to de-alias the images via multi-scale downsampling and synthesis paths with skip connections. 
To enforce temporal consistency between frames of cardiac cine sequences, the U-net can be equipped with either 3D convolutions to perform on the entire image sequences ($x$-$y$-$t$) \cite{hauptmann2019real} or 2D convolutions on the 2D spatio-temporal domain ($x$-$t$) for more efficient training \cite{kofler2019spatio}. These models \cite{hauptmann2019real,kofler2019spatio} showed better image quality and faster reconstruction speed recovered from undersampled 2D golden-angle radial cine CMR when compared to CS image reconstruction techniques. 

Since these methods solely work on information from the initial zero-filling image data and information extracted from training samples, they can also be classified as purely data-driven approaches. Though these methods are simple and efficient in reconstructing images, one significant drawback is that they cannot guarantee the consistency of the reconstructed image with respect to the measured $k$-space data. This limits their performance in comparison to the  model-driven approaches which will be discussed in the following. 

\subsection{Model-driven DL for CMR Reconstruction}
Another type of approaches for CMR image reconstruction are model-driven DL techniques based on unrolled schemes inspired by conventional iterative CS reconstruction, which also represents the majority of the work in this field. These methods utilise the physical domain knowledge with the incorporation of raw $k$-space data, combining the power of data-driven learning with the physics-derived model-based framework (Eq. \ref{eq1:problem_fomulation}). In contrast to CS-based methods where regularisation terms need to be specified manually beforehand, DL-based reconstruction approaches implicitly learn such prior information and regularisation from the routinely performed MR data.

In detail, model-driven DL-based solutions \cite{schlemper2017deep,schlemper2018deep,qin2019convolutional,aggarwal2018modl,hammernik2018learning,duan2019vs} propose to learn the unrolled optimisation and embed the iterative reconstruction process in a learning setting based on Eq. \ref{eq:alternate_minimisation}. Specifically, Eq. \ref{eq:proximal_operator} is commonly modelled via leveraging deep neural networks (DNNs) to learn the prior knowledge from the training data itself \cite{schlemper2018deep,qin2019convolutional,aggarwal2018modl}. This allows the model to fully exploit the information contained in the population data for the regularisation and alleviates the need to explicitly pre-define them based on certain assumptions. Accordingly, Eq. \ref{eq:data_fidelity} can be derived as close-form solutions, which then can be embedded as DC layers in DNNs via exact update for DC \cite{schlemper2018deep,qin2019convolutional,duan2019vs} or through numerical optimisation such as conjugate gradient updates \cite{aggarwal2018modl,biswas2019dynamic,mardani2018deep}, depending on the way to solve Eq. \ref{eq:data_fidelity}. Therefore, model-driven approaches based on Eq. \ref{eq:alternate_minimisation} can be generally represented as the compact form:
\begin{subequations}
\begin{align}
{\mathbf{z}}^{(i+1)} &=  \textnormal{DNN}({\bf{x}}^{(i)}), \label{eq:DNN}
\\
{\bf{x}}^{(i+1)} &= \textnormal{DC}({\bf{z}}^{(i+1)}; \mathbf{y}, \mathbf{E}, \lambda, \mu), 
\end{align}
\end{subequations} 
which alternates the reconstruction process between image de-aliasing and DC steps for a pre-defined number of iterations. This represents the most commonly used scheme for DL-based CMR reconstruction. For other types of model-driven approaches (including applications not limited to CMR), please refer to \cite{hammernik2020machine} for details.

Based on types of deep-learned regularisation priors, we generally group these unrolled DL techniques into two classes: image-domain-regularisation and complementary-regularisation approaches. More specific network architectures and applied scenarios in cardiac imaging are introduced hereafter.  

\subsubsection{Learning Unrolled Optimisation with Image Domain Regularisation}

\paragraph{Unrolled Convolutional Neural Networks}
Typical approaches for constructing Eq. \ref{eq:DNN} are to employ convolutional neural networks (CNNs) as regularisation operators in image domain. State-of-the-art methods such as DC-CNN \cite{schlemper2017deep,schlemper2018deep} and MoDL \cite{aggarwal2018modl,biswas2019dynamic} used 5 layers of CNN with a residual connection for 10 iterations. To exploit spatio-temporal redundancies, Schlemper et al. \cite{schlemper2018deep} proposed to incorporate a data sharing (DS) layer in a 3D ($x$-$y$-$t$) CNN cascade network, which allows to utilise similar information contained in neighboring $k$-space samples. The method was performed on retrospectively undersampled single-coil cardiac cine MR images, where the single-coil setting enables solution to Eq. \ref{eq:data_fidelity} to be analytically computed as an exact update for DC step. For multi-coil acquisition, variational network (VN) \cite{hammernik2018learning} proposed to embed a combined PI and CS reconstruction in a DL-based unrolled gradient descent scheme. An extension of it has also been applied for dynamic cine CMR via learning the complex spatio-temporal convolutions corresponding to spatio-temporal total variation regularisation \cite{hammernik2019dynamic}. Beyond applications on 2D dynamic cine CMR, Fuin et al. \cite{fuin2019variational} has extended VN to a multi-scale VN architecture for fast reconstruction of undersampled motion-compensated free-breathing whole-heart 3D ($x$-$y$-$z$) CMRA, and CINENet \cite{kustner2020cinenet} was introduced for highly prospectively undersampled 3D Cartesian cardiac cine data via exploiting spatio-temporal redundancies by cascaded and complex-valued (3+1)D spatial and temporal convolutions.

\begin{figure*}[!t]
\centering
\includegraphics[width=\linewidth]{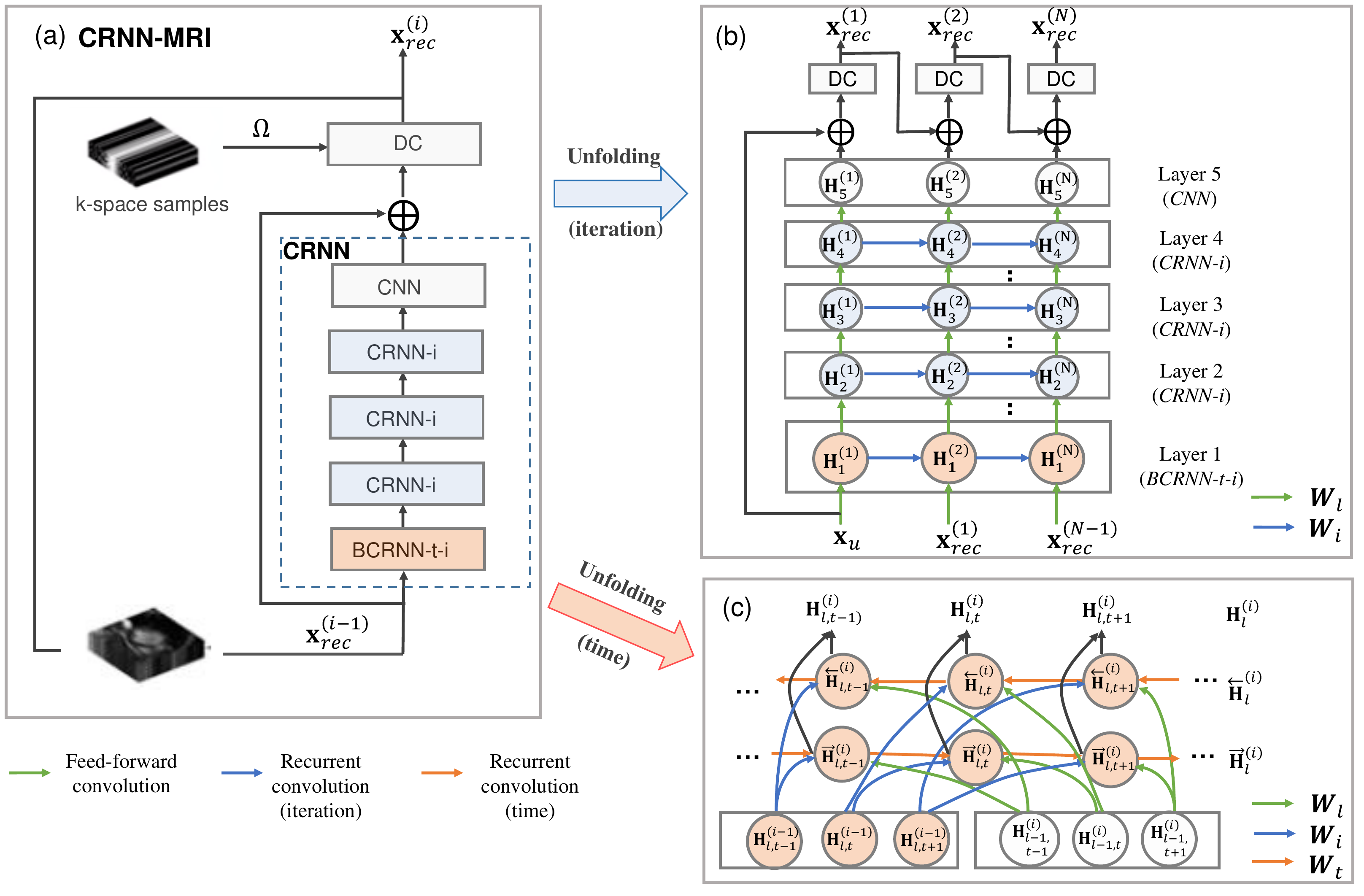}
\caption{(a) Architecture of CRNN-MRI network \cite{qin2019convolutional} for CMR image reconstruction. (b) The structure of CRNN-MRI when unfolded over iterative reconstruction steps. (c) The structure of bidirectional RNN layer (BCRNN-t-i) when unfolded over the time sequence. The method is able to jointly exploit spatio-temporal redundancies and iteration dependencies in a recurrent fashion with a single framework.}
\label{fig:CRNN}
\end{figure*}

\paragraph{Unrolled Convolutional Recurrent Neural Networks}
Another type of representative approach for learning the deep priors in Eq. \ref{eq:DNN} is via recurrent architectures. Recurrent neural networks (RNNs) are known to be effective in processing sequential information with an internal state acting as ``memory" and thus have been widely employed for modelling the iterative reconstruction steps \cite{qin2019convolutional,zhou2020dudornet,huang2020dynamic,wang2019pyramid}. As one of the pioneering work in this direction, CRNN-MRI \cite{qin2019convolutional} proposed to embed the iterative optimisation scheme (Eq. \ref{eq:alternate_minimisation}) into a recurrent network structure with hidden-to-hidden connections between each iteration step, which allows effective information propagation along iterations and avoids redundant computations. Furthermore, to fully exploit spatio-temporal redundancies, bidirectional RNNs were also designed to model the sequential dependencies along temporal axis together with recurrence over iteration dimension. An illustrative diagram of the model is shown in Fig. \ref{fig:CRNN}, where unfolded structures of each layer are given with more details. The method was validated on retrospectively undersampled single-coil cardiac cine MRI, which achieved better and more efficient reconstructions compared to CNN architectures with fewer number of network parameters. 

\paragraph{Motion Compensated Dynamic CMR Reconstruction}
Due to the presence of motion patterns in dynamic CMR acquisition, combining motion estimation and motion compensation with model-driven DL-based reconstruction methods has also been explored. End-to-end motion compensated reconstruction framework \cite{huang2020dynamic,huang2019dynamic,seegoolam2019exploiting} commonly consists of DL-based motion estimation components \cite{qin2018joint} estimating motion information between frames within undersampled MRI sequences \cite{seegoolam2019exploiting} or with respect to fully sampled reference frames \cite{huang2020dynamic}. The estimated motion then can be exploited to generate motion compensated images to be infused into reconstruction process \cite{huang2019dynamic,huang2020dynamic} or to propagate $k$-space data along the full temporal dynamics via data consistency \cite{seegoolam2019exploiting} for better reconstruction quality. These approaches have been validated on single-coil 2D cardiac cine MRI data with retrospective undersampling, and incorporating motion information into reconstruction process has shown to be effective in reconstructing aggressively accelerated data \cite{seegoolam2019exploiting}.

\subsubsection{Learning Unrolled Optimisation with Complementary Regularisation}
\begin{figure*}[!t]
\centering
\includegraphics[width=\linewidth]{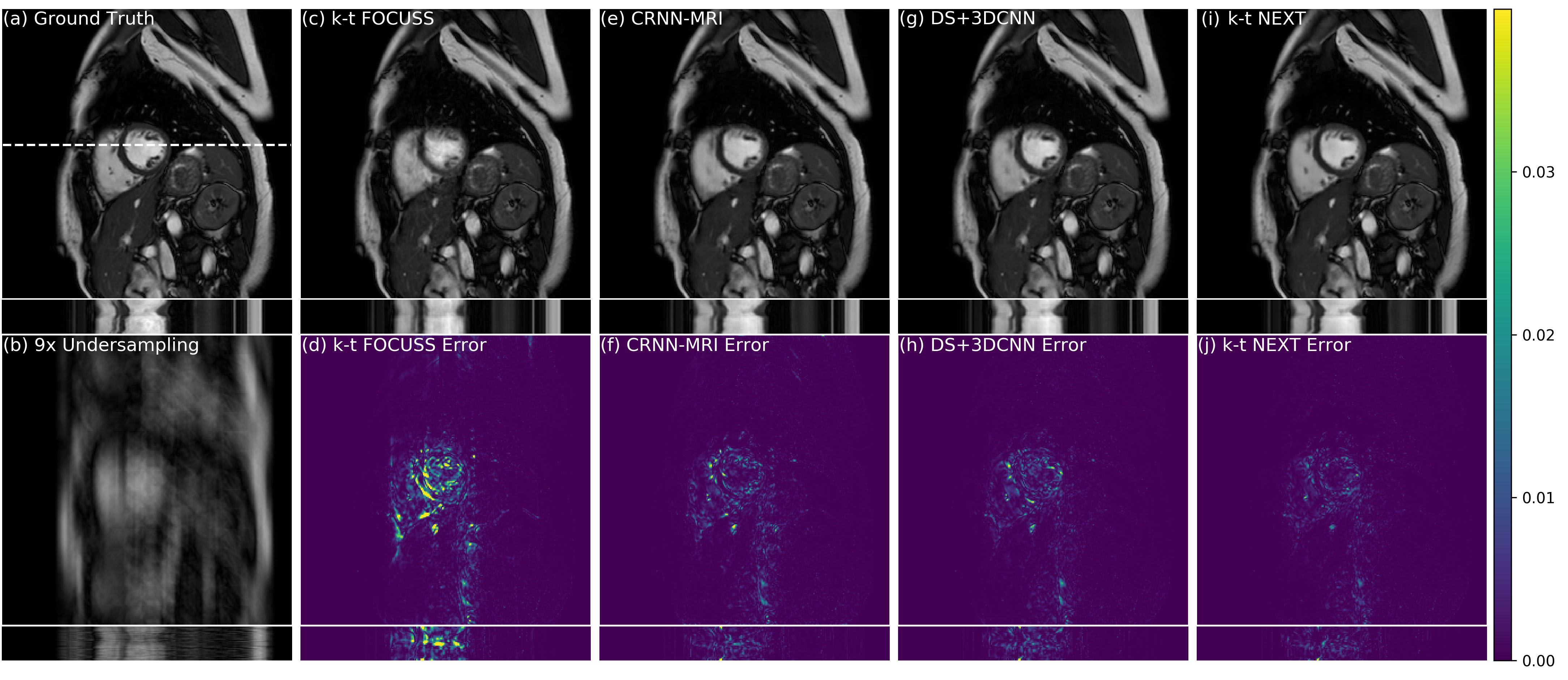}
\caption{CMR image reconstruction for 9-fold accelerated single-coil data along with their error maps.
(a) Fully sampled reference (b) $9 \times$ undersampled image (c, d) $k$-$t$ FOCUSS \cite{jung2007improved} (e, f) CRNN-MRI \cite{qin2019convolutional} (g, h) DS+3DCNN \cite{schlemper2018deep} (i, j) $k$-$t$ NEXT \cite{qin2019k}. Model-driven DL approaches outperform CS-based $k$-$t$ FOCUSS. Learning with complementary regularisation ($k$-$t$ NEXT) achieves better reconstruction than methods with single domain regularisation.}
\label{fig:xy}
\end{figure*}

Similar to conventional CS-based approaches, model-driven DL for CMR reconstruction can also be learnt with complementary regularisation. Typical $k$-$t$ methods \cite{lustig2006kt,jung2007improved,otazo2010combination} exploit signal sparsity and regularise the data in the combined spatial and temporal Fourier ($x$-$f$) domain for dynamic CMR reconstruction. This can also be similarly leveraged with DL by learning deep priors in complementary image and $x$-$f$ spaces. Qin et al. \cite{qin2019k} proposed the $k$-$t$ NEXT approach along this direction, which alternates the learned DL-based regularisation between both image and $x$-$f$ domains in an iterative fashion for single-coil dynamic CMR reconstruction. Comparison results of $k$-$t$ NEXT against other model-driven DL approaches and CS-based method are shown in Fig. \ref{fig:xy}, where it can be seen that exploiting complementary domain knowledge can achieve better performance than single image-domain reconstructions, with better preservation of fine details and sharp structures. An extension of it has also been investigated for multi-coil dynamic CMR data with a complementary time-frequency domain network based on variable splitting technique in PI setting \cite{qin2020complementary}. 
In addition to the $x$-$f$ regularisation, MoDL-SToRM \cite{biswas2019dynamic} combines MoDL reconstruction \cite{aggarwal2018modl} along with a complementary image regularisation penalty, i.e., SmooThness regularisation on manifolds (SToRM) prior \cite{poddar2015dynamic}, to recover free-breathing cardiac cine MR images. Similarly, this framework alternates between a learned regularisation of the image using CNN, an analytically defined SToRM prior, and a conjugate gradient DC step for multi-channel measurements. Besides, the complementary information can also be exploited in both $k$-space and image space, and this has been shown in \cite{wang2019dimension} where the frequency domain network and image domain network were combined and trained with a multi‐supervised technique for the single-coil 2D cardiac cine MR image reconstruction.
All these above methods have demonstrated the benefits of combining complementary regularisation, achieving better performance than single image-domain approaches.

\subsection{DL for K-space based CMR Reconstruction}
An alternative class of techniques for MRI reconstruction is based on $k$-space interpolation. This is represented by methods such as GRAPPA (generalized autocalibrating partially parallel acquisitions) \cite{griswold2002generalized} which recovers the missing data directly in $k$-space by employing autocalibrating signal (ACS) estimated for each subject. Similar to this, a CNN-based approach has also been proposed to improve the $k$-space interpolation for PI-based MRI reconstruction. The RAKI (robust artificial-neural-networks for $k$-space interpolation) reconstruction approach \cite{akccakaya2019scan} enables non-linear estimation of missing $k$-space lines from acquired multi-coil data via using CNNs trained on ACS data for $k$-space interpolation. This has been applied for reconstruction of retrospectively undersampled myocardial 2D T1 mapping data with improved noise resilience. Note, that RAKI is a scan-specific and database-free DL approach, which does not require training datasets but relies on ACS data and separate CNN training for each scan.

\subsection{Network Learning}
In general, most of the aforementioned supervised DL methods rely on large databases of training data to learn the population-based prior information and the mapping between pairs of undersampled data and reference MRI scans. Due to the limited availability of the acquired images, most approaches consider retrospectively simulating the undersampled data from fully-sampled raw $k$-space data, which are then used as input and target for DL models during the training process. To train these supervised models, metrics that measure the similarity between the reconstructed image $\mathbf{x}$ and the reference image $\mathbf{x}_\text{ref}$ are used as loss functions. One commonly used loss function is the mean-squared-error (MSE) which measures the L2 distance between $\mathbf{x}$ and $\mathbf{x}_\text{ref}$ in the training set $\Omega$:
\begin{equation}
    \mathcal{L}_\text{mse}(\theta)= \frac{1}{n_\Omega}\sum\limits_{(\mathbf{x}, \mathbf{x}_\text{ref}) \in \Omega} {\left\| {{{\mathbf{x}}_\text{ref}} - {{\mathbf{x}}}} \right\|_2^2},
\end{equation}
Here $\theta$ is the network parameters and ${n_\Omega}$ is the number of training samples. Though the MSE metric is easy to optimise, it often tends to generate blurry reconstructions due to the averaging effect. To remedy this, the L1 distance or structural similarity index measure (SSIM) are also often employed as the loss function to improve upon the sharpness of structures, such that:
\begin{equation}
        \mathcal{L}_\text{ssim}(\theta)= \frac{1}{n_\Omega}\sum\limits_{(\mathbf{x}, \mathbf{x}_\text{ref}) \in \Omega} \left(1-\text{SSIM}(\mathbf{x},\mathbf{x}_\text{ref}) \right).
\end{equation}
Additionally, adversarial loss functions \cite{goodfellow2014generative} have also gained in popularity for reconstructing images with improved perceptual quality. Such a loss function has also been used for MRI reconstruction \cite{yang2017dagan,seitzer2018adversarial,quan2018compressed}. The adversarial approach works by learning a discriminator to discriminate whether a given sample is from a reconstruction network (generator) or from the reference distribution. The adversarial loss has shown to be effective in generating images with sharper textures and finer details compared to MSE. However, it is known to be unstable in training and is also prone to hallucinate unrealistic artefacts. This is problematic for MRI reconstruction in preserving the data fidelity and anatomical structures. Therefore, the adversarial loss is often combined with the content losses such as MSE and L1 norm to inform the reconstruction with image contents. An alternative way to balance the data fidelity and perceptual quality uses deep network interpolation \cite{qin2020deep}, where these two effects can be trade off with a interpolation coefficient in parameter space. Finally, perceptual loss terms \cite{ledig2017photo} have also been introduced to improve the reconstruction details based on feature representations of the VGG network \cite{simonyan2014very}, and this has also been applied for CMR reconstruction \cite{seitzer2018adversarial}.

\section{Discussion}
Most of the current research shows exciting and promising developments of AI algorithms in image reconstruction for cardiac imaging. These are mostly attributed to their strong ability in recovering high quality images with fast reconstruction speed, which therefore can further increase the acceleration capability. Current DL-based approaches, especially models learning an unrolled optimisation \cite{schlemper2018deep,qin2019convolutional,wang2019dimension,biswas2019dynamic,fuin2019variational,kustner2020cinenet}, have shown to outperform conventional CS-based methods for CMR reconstruction, and their application in PI also indicates their potentiality in facilitating fast single-breath-hold 2D cardiac cine imaging. Though promising results have been achieved in recent developments, there are still some challenges yet to be solved for their use in clinical applications. 
One of the limitations of current studies for CMR image reconstruction is their limited validation in clinical practice. Most of the existing DL-based works focus on reconstructing images with retrospectively undersampling based on models trained on simulated undersampled data. Though such retrospective study can provide useful insights and initial validation of employed methods, it still remains unclear how well these methods can perform on prospectively undersampled data, which is a more clinically practical scenario. In addition, majority of current DL research for dynamic CMR image reconstruction works on simulated single-coil acquisition setting, whereas only a handful of works propose to address the more practical and commonly used multi-coil acquisitions. This is mainly due to the increased computational and model complexities introduced by the multi-coil data as well as the extra temporal dimension. Thus, efficient DL models for PI-based dynamic CMR image reconstruction are also highly desirable. Besides, as most work investigates reconstruction with Cartesian undersampling patterns, efforts towards non-uniform undersampling strategies such as radial sampling and golden angle sampling are also necessary, which represent the commonly used sampling strategies in acceleration of 2D cardiac MR imaging in practice.

Another issue with DL-based CMR image reconstruction methods is their generalisation ability when presented with data that looks different to the data used in training. It is not yet systematically evaluated how well DL models trained on MR scans with specific scanners, pulse sequences, imaging resolutions or acceleration factors can generalise to other MR scans with different acquisition parameters. Furthermore, the generalisation ability of these models on unseen pathologies still needs to be explored in detail. For instance, it will be interesting and useful to see if models trained on healthy cohorts can faithfully recover the pathological patterns of patients with cardiovascular diseases. Though a recent study \cite{qin2020complementary} has made efforts towards this direction and shows promising generalisation potential of the DL-based approach, more thorough and systematic evaluations are still needed for a wider validation on this aspect including radiologists' evaluation to improve their translation into clinical practice.

Additionally, in comparison to conventional optimisation-based CS methods, training DL models normally requires large databases of training data. However, such large training database is not available for most of the CMR image reconstruction studies, and currently most existing works rely on a small set of training samples with a great deal of data augmentations to remedy this. Though such strategy has shown to be effective, it could still limit the performance potential of such DL-models as well as their validations due to the limited amount of available data. The FastMRI datasets \cite{knoll2020fastmri} provide large databases of  knee ($\sim$1000) and brain ($\sim$7000) MR images together with the corresponding k-space data, which brings a great opportunity for accelerating MR imaging with AI. However, such a large archive of images and their corresponding raw k-space data is still lacking in cardiac imaging, and efforts towards building a large CMR database will be helpful to further push the developments of the use of AI in this field. Besides, methods beyond supervised learning such as self-supervised learning and unsupervised learning will also be interesting to be explored to reduce the expensive need of fully sampled reference data for training, and models in transfer learning and domain adaptation can also be investigated to effectively exploit the large amount of available data in other domains for CMR reconstruction. 

\section{Conclusion}
Based on recent advancements in AI technology, the field of cardiac image reconstruction also sees its exciting developments with the use of AI algorithms in recent years. This chapter provides an overview of these developments with a major focus on introducing the existing popular DL-based approaches for CMR image reconstruction as well as the current challenges. Though DL approaches for CMR image reconstruction are not yet established for clinical applicability and will require more thorough evaluation, they have already shown a huge potential for accelerating and improving the CMR image acquisition and reconstruction workflow.
\bibliographystyle{spmpsci}
\bibliography{references}
\end{document}